\documentclass[aps,prapplied,reprint,groupedaddress,amsmath,amssymb]{revtex4-2}

\usepackage{esint}
\usepackage{graphicx}
\usepackage{dcolumn}
\usepackage{bm}
\usepackage{float}
\usepackage{upgreek}
\usepackage{array}
\usepackage{soul}
\usepackage{color}

\usepackage[colorlinks, linkcolor=blue, citecolor=blue, urlcolor=blue,breaklinks=true]{hyperref}

\newcolumntype{L}[1]{>{\raggedright\let\newline\\\arraybackslash\hspace{0pt}}m{#1}}
\newcolumntype{C}[1]{>{\centering\let\newline\\\arraybackslash\hspace{0pt}}m{#1}}
\newcolumntype{R}[1]{>{\raggedleft\let\newline\\\arraybackslash\hspace{0pt}}m{#1}}

\begin{document}

\title{Compact vacuum gap transmon qubits: \\Selective and sensitive probes for superconductor surface losses}

\author{M. Zemlicka}
\email{martin.zemlicka@ist.ac.at}
\affiliation{Institute of Science and Technology Austria, 3400 Klosterneuburg, Austria}
\author{E. Redchenko}
\affiliation{Institute of Science and Technology Austria, 3400 Klosterneuburg, Austria}
\author{M. Peruzzo}
\affiliation{Institute of Science and Technology Austria, 3400 Klosterneuburg, Austria}
\author{F. Hassani}
\affiliation{Institute of Science and Technology Austria, 3400 Klosterneuburg, Austria}
\author{A. Trioni}
\affiliation{Institute of Science and Technology Austria, 3400 Klosterneuburg, Austria}
\author{S. Barzanjeh}
\email{Present address: Institute for Quantum Science and Technology (IQST), University of Calgary, Calgary, AB, Canada.}
\affiliation{Institute of Science and Technology Austria, 3400 Klosterneuburg, Austria}
\author{J. M. Fink}
\email{jfink@ist.ac.at}
\affiliation{Institute of Science and Technology Austria, 3400 Klosterneuburg, Austria}

\date{\today}

\begin{abstract}
State-of-the-art transmon qubits rely on large capacitors which systematically improves their coherence due to reduced surface loss participation. However, this approach increases both the footprint and the parasitic cross-coupling and is ultimately limited by radiation losses - a potential roadblock for scaling up quantum processors to millions of qubits. 
In this work we present transmon qubits with sizes as low as 36$\,\times\,$39$\,\upmu$m$^2$ with $\gtrsim$100\,nm wide vacuum gap capacitors that are micro-machined from commercial silicon-on-insulator wafers and shadow evaporated with aluminum. 
We achieve a vacuum participation ratio up to 99.6\% in an in-plane design that is compatible with standard coplanar circuits. Qubit relaxation time measurements for small gaps with high vacuum electric fields of up to 22\,V/m reveal a double exponential decay indicating comparably strong coupling to long-lived two-level-systems. The exceptionally high selectivity of $>$20\,dB to the superconductor-vacuum surface allows to precisely back out the sub-single-photon dielectric loss tangent of aluminum oxide exposed to ambient conditions. In terms of future scaling potential we achieve a qubit quality factor by footprint area of $20\,\upmu \mathrm{s}^{-2}$, which is on par with the highest $T_1$ devices relying on larger geometries and expected to improve substantially for lower surface loss superconductors. 
\end{abstract}

\maketitle

\section{Introduction\label{intro}}
One of the most promising ways towards a scalable quantum computer using quantum error correction \cite{barends2014,corcoles2015}, is to utilize superconducting transmon qubits \cite{koch2007}. The transmon is formed by a Josephson junction (JJ) shunted with a large capacitance that reduces its sensitivity to charge noise. State-of-the-art coherence qubits realize this capacitance with large footprint capacitor plates and cavities \cite{paik2011,lei2020}. 
This is because the electric field per excitation is decreased and distributed over a larger volume, thus lowering the effective coupling to impurities such as parasitic two-level systems (TLS) \cite{phillips1987} that are 
localized in the material
interfaces: metal-substrate (MS), substrate-air (SA), and metal-air (MA).
However, this approach starts to be limited by bulk losses of the best know substrates \cite{read2022}, it increases the physical size of the qubit on the chip up to $\sim$mm$^2$
which not only limits the integration density but also increases the parasitic cross coupling 
between circuit elements.
Another disadvantage is that large transmons form an antenna 
which not only leads to increased radiation loss \cite{sage2011,rafferty2021} but also to increased stray photon absorption with the associated breaking of Cooper pairs \cite{pan2022}.
The same shunt capacitance and qubit energy spectrum 
can in principle also be achieved if the capacitor electrodes and gaps are scaled down to much smaller dimensions
but this rapidly increases the dielectric losses due to 
larger electric fields that are more localized in the lossy surfaces.

Dielectric losses, e.g.~due to resonant TLS absorption, are in fact the predominant factor limiting energy relaxation times and in turn also the achieved coherence times in fixed frequency transmons in the GHz range. This strongly motivates a detailed geometrical analysis \cite{gambetta2017,martinis2022} and design optimization to speed up the development of longer lifetime qubit designs with higher integration density \cite{siddiqi2021}. 
Such optimizations have to rely on well known surface properties and a considerable number of studies were investigating these loss channels \citep{barends2010, sage2011, wenner2011, sandberg2012, wang2015, chu2016, woods2019}. Nevertheless, it has been difficult to reliably distinguish the loss contributions from different interfaces.
In some of the studies it was pointed out that MS and SA are the lossiest interfaces \cite{barends2010,wenner2011} and also that the loss from the MS interface remains unexplained \cite{megrant2012}. Some studies suggest that the losses at the MS interface could be intrinsic \cite{macha2010,herman2016}. The MA interface could to some degree be controlled with metal surface cleaning techniques and improved by using superconducting materials with less parasitic oxide such as TiN or NbTiN \cite{barends2010, sage2011}. There were attempts of reducing the surface sensitivity of the electric field by trenching gaps in coplanar waveguide (CPW) resonators \cite{calusine2018}, but even after very deep etching there is still about 50\% of the electric field stored in the silicon substrate due to the large dielectric permittivity.

Vacuum gap capacitors \cite{cicak2009} have the majority of the electric field energy stored in the lossless vacuum and ideally they are solely sensitive to the MA interface losses. 
A circuit using such a capacitor
can thus serve as a very sensitive probe to provide detailed information about the superconductor surface properties. Previously investigated vacuum gap capacitor circuits \cite{cicak2009, cicak2010, bosman2017} utilize a drum capacitor 
that is arranged out-of-plane and relies on a plasma-assisted chemical release step. The resulting capacitor gap is very sensitive to mechanical instabilities and the temperature and deposition dependent stress in the thin film. Furthermore, the fabrication method relies on a sacrificial layer with potential residues after removal which can modify the quality factor and the extracted surface loss tangent. 

Here we utilize commercial, smart-cut, high resistivity silicon-on-insulator (SOI) wafers, where a 220\,nm thick silicon membrane is suspended on a silicon handle wafer by a $3\,\upmu$m thick silicon dioxide layer as a circuit substrate \cite{dieterle2016, keller2017}. Using silicon plasma etching, angled e-beam aluminum evaporation \cite{pitanti2015} and dry hydrogen fluoride (HF) vapor release of the oxide below the membrane yields vacuum gap finger capacitors with close to all
of the electric field energy stored in vacuum and a surface loss participation of the MA interface that is more than 2 orders of magnitude higher than the simulated MS, SA (or bulk) contributions. 
Moreover, the electric field in the capacitor is very confined with an effective relative permittivity of $\approx 1$, which prevents parasitic coupling and radiation. The in-plane design is reproducible, structurally reliable and compatible with common coupling techniques. 

To quantify the loss properties for different vacuum gaps ranging from 0.1 to 1$\,\upmu$m we connect them in parallel either with a meander inductor to form a lumped element $LC$ resonator \cite{geerlings2012} or with a JJ to form a transmon qubit with a lumped element readout resonator \cite{suri2013}. Measuring the resonator internal quality factor $Q_{\mathrm{R}}$ and the qubit lifetime $T_1$ we obtain detailed information on how the commonly used aluminum oxide surface limits the superconducting circuit's energy dissipation.

\begin{figure}[t]
\includegraphics[width=1.0\columnwidth]{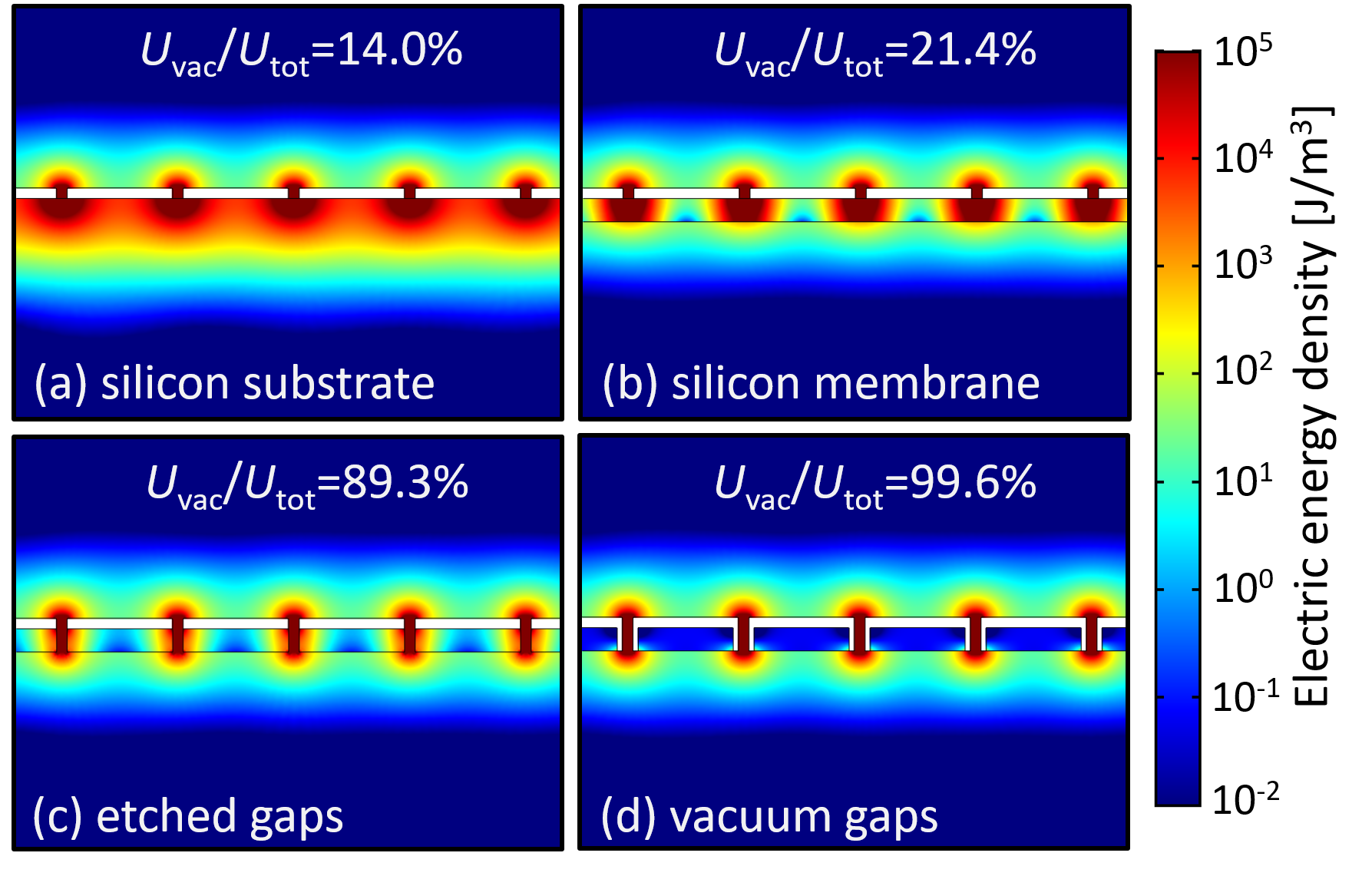}
\caption{\label{fig:density} 
\textbf{Electric field energy density}. 
Finite element method simulations show how the electric field is distributed between vacuum and dielectrics for 1\,$\upmu$m wide metal fingers (white) separated by 100\,nm wide gaps situated on (a) a standard silicon substrate, (b) a 220\,nm thin silicon membrane, (c) dry-etched suspended silicon beams, and (d) beams with side wall metalization forming a vacuum-gap capacitor.}
\end{figure}

\begin{figure}[t]
\includegraphics[width=\columnwidth]{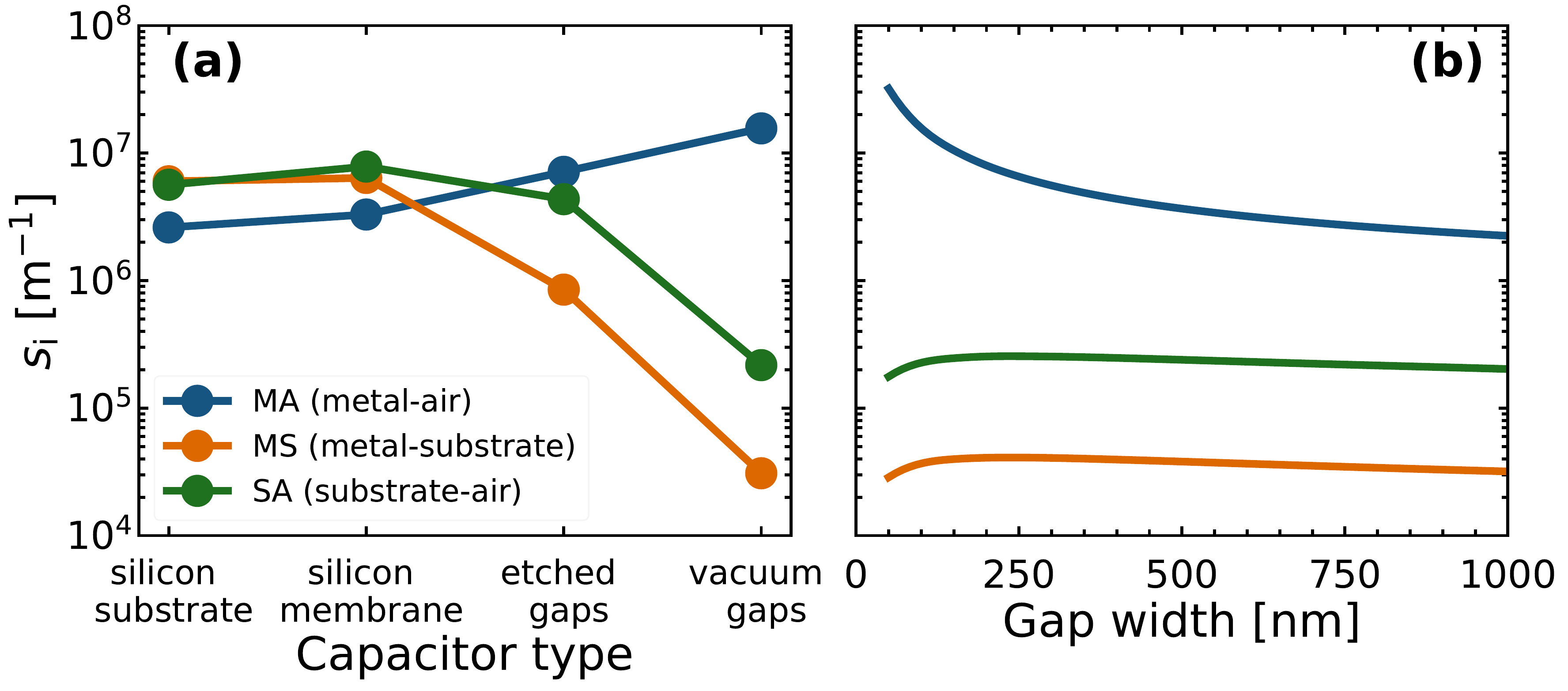}
\caption{\label{fig:sens} 
\textbf{Electric field surface sensitivity.} 
COMSOL simulation of the surface sensitivity to the electric field calculated for the interfaces: metal-air (MA), metal-substrate (MS) and substrate-air (SA) for 
(a) different type of capacitors with 100\,nm gap width, and 
(b) the superconducting vacuum gap capacitor with variable gap width (circles). Connecting lines serve as a guide to the eye.}
\end{figure}

\section{Simulation}
To predict the electric properties of the proposed capacitor we use the ACDC module of COMSOL's Multiphysics simulation software \cite{comsol}. In Fig.~\ref{fig:density} we see how replacing a bulk silicon substrate with a nano-membrane, removing the dielectric from the gaps, and metalizing the sidewalls of the suspended dielectric beams subsequently increases the ratio of electric field energy stored in vacuum from 14\% to 99.6\%. The remaining 0.4\% are stored in the bulk high resistivity silicon 
with a low loss tangent of typically $\tan{\delta}\approx 10^{-7} - 10^{-9}$, that does not limit the dissipation in this circuit. 

The predominant source of energy relaxation is associated with 
impurities or TLS with electric dipoles that are located in material interfaces and electrically coupled to the circuit capacitor. To quantify how much each surface interface is involved in this effect we define the surface sensitivity to the electric field $s_i=\frac{p_i}{\varepsilon_it_i}$, where $p_i=\frac{1}{U_{\mathrm{tot}}}\int\displaylimits_{V_i}\frac{\vec{E}\times \vec{D}}{2}\mathrm{d}V$ is the participation ratio \cite{wang2015} of the electric field energy stored in the parasitic layer on the particular interface with volume $V_i$, thickness $t_i$, dielectric constant $\varepsilon_i$ and $U_{\mathrm{tot}}$ the total energy. Assuming that the electric and displacement fields $\vec{E}$ and $\vec{D}$ are constant across the parasitic layer thickness and that the relative permittivity $\varepsilon_i$ is uniform across the whole layer we can express $s_i$ defined only by the capacitor geometry
\begin{eqnarray} \label{eq:sens}
s_i=\frac{1}{U_{\mathrm{tot}}}\int\displaylimits_{S_i}\frac{\varepsilon_0|\vec{E}|^2}{2}\mathrm{d}S,
\end{eqnarray}
where $S_i$ is the total surface of the interface and the electric field surface distribution can be extracted from the simulation. 

Figure~\ref{fig:sens}(a) shows the simulated surface sensitivity of each interface for the same type of capacitors as for the energy density plot in Fig.~\ref{fig:density}. Removing the dielectric and focusing the electric field into the vacuum rapidly decreases $s_{\mathrm{SA}}$, $s_{\mathrm{MS}}$ and increases $s_{\mathrm{MA}}$ leading to a two orders of magnitude difference. 
Using the simulated values and the assumed properties of the parasitic interface layers $t_\mathrm{i}=3$\,nm and $\varepsilon_\mathrm{i}=10$ \cite{wang2015} for a 100\,nm gap width we evaluate that the MA interface has by far the largest contribution of $p_{\mathrm{MA}}\approx3\%$ of the total energy, which was previously associated with the vacuum participation.
The contributions from other dielectric interfaces $p_{\mathrm{MS}}\approx0.012\%$, $p_{\mathrm{SA}}\approx0.075\%$ and from the bulk silicon $p_{\mathrm{Si}}\approx0.4\%$ are negligible in comparison with $p_{\mathrm{MA}}$. 
Figure~\ref{fig:sens}(b) shows that this is most pronounced
when the vacuum gap width is $\lesssim$ membrane thickness of 220\,nm.

\section{Fabrication\label{fab}}
All the structures are patterned using electron beam lithography and standard lift-off and silicon dry etching methods as discussed in detail in Appendix~A and as shown in Fig.~\ref{fig:sem}. After etching of the capacitor gaps into the SOI device layer, the vacuum gap is formed by angle evaporation of aluminum covering both the top and sidewalls of the partially suspended silicon beams in a single evaporation. Subsequently the JJ is fabricated with the standard Dolan bridge technique and a patch layer involving ion gun etching of the aluminum contact surface prior to the evaporation is used to yield reliable contact. 
In the last step the membrane holding the qubit and LC resonator is released in the entire region of the qubit/resonator circuit by fully etching the 3\,$\upmu$m silicon dioxide buffer layer separating the device layer from the handle wafer with dry HF vapor. A finished small qubit device is shown on various scales in Fig.~\ref{fig:sem}. The qubit-resonator coupling design was implemented with a comparably large non-trenched capacitor to avoid its potential impact on the qubit loss characterization.
In the future this part could also be realized with a small vacuum gap.

\section{Results}
\subsection{Resonator measurements}
\begin{figure}[t]
\includegraphics[width=1.0\columnwidth]{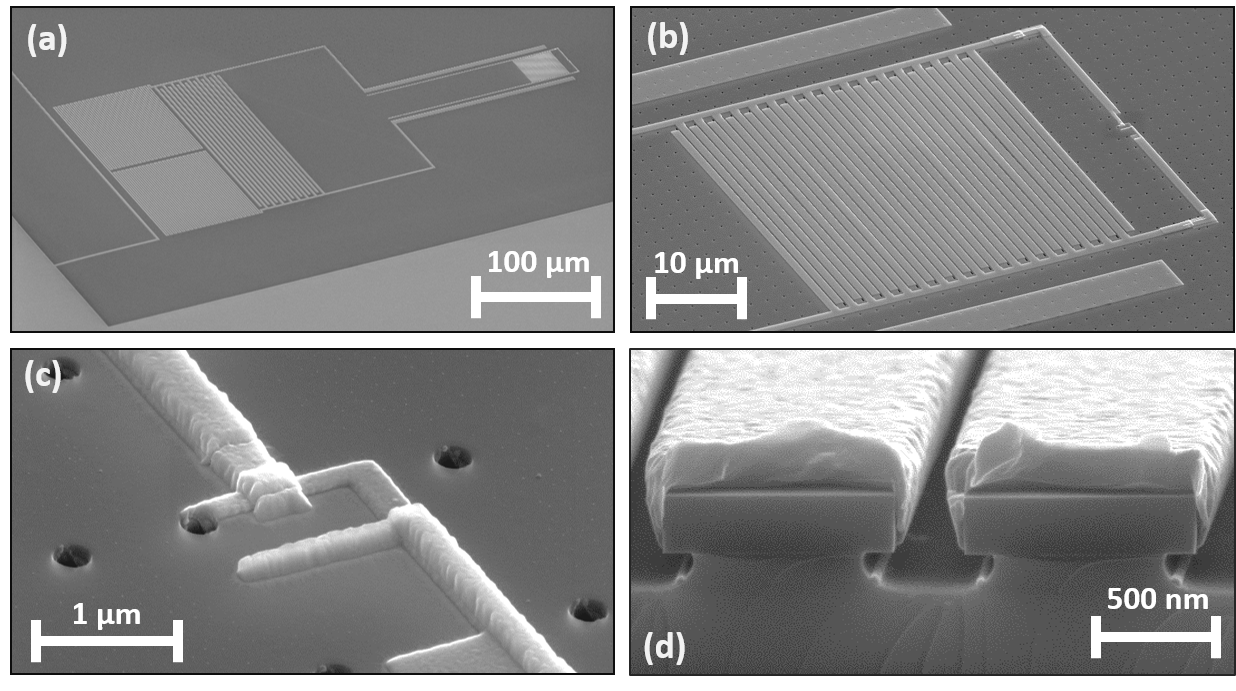}
\caption{\label{fig:sem} \textbf{Scanning electron microscope images of a fabricated qubit device.} 
(a) Meander inductor $LC$ resonator capacitively coupled to the transmon qubit and inductively coupled to a shorted coplanar transmission line. 
(b) Enlarged view of the qubit (a vacuum-gap finger capacitor shunted with a JJ).
(c) Enlarged view of the shadow evaporated JJ. Also visible are dry etched holes in the silicon device layer that facilitate the HF vapor release of the membrane.
(d) Cross-section of the vacuum gap capacitor obtained by cleaving and imaging a sacrificial device before the HF vapor removal of the silicon dioxide below the beams.}
\end{figure}

To quantify the dielectric microwave losses we first fabricate lumped element resonators using a vacuum gap capacitor with $C \approx 50\,$fF with different gap size and footprint connected in parallel to a meander inductor with $L\approx 5\,$nH similar to the one shown in Fig.~\ref{fig:sem} but without the qubit, see also Appendix A. 
We then measure the frequency response of the reflection scattering paramter
around the fundamental mode resonance frequency $\omega_\text{R}/(2\pi)\approx 8.5$\,GHz using a vector network analyzer (VNA) with varying probe power. Fitting both scattering parameter quadratures with the model for single port resonant coupling 
we obtain the internal quality factor $Q_\mathrm{R}$ of the resonator.

\begin{figure}[t]
\includegraphics[width=\columnwidth]{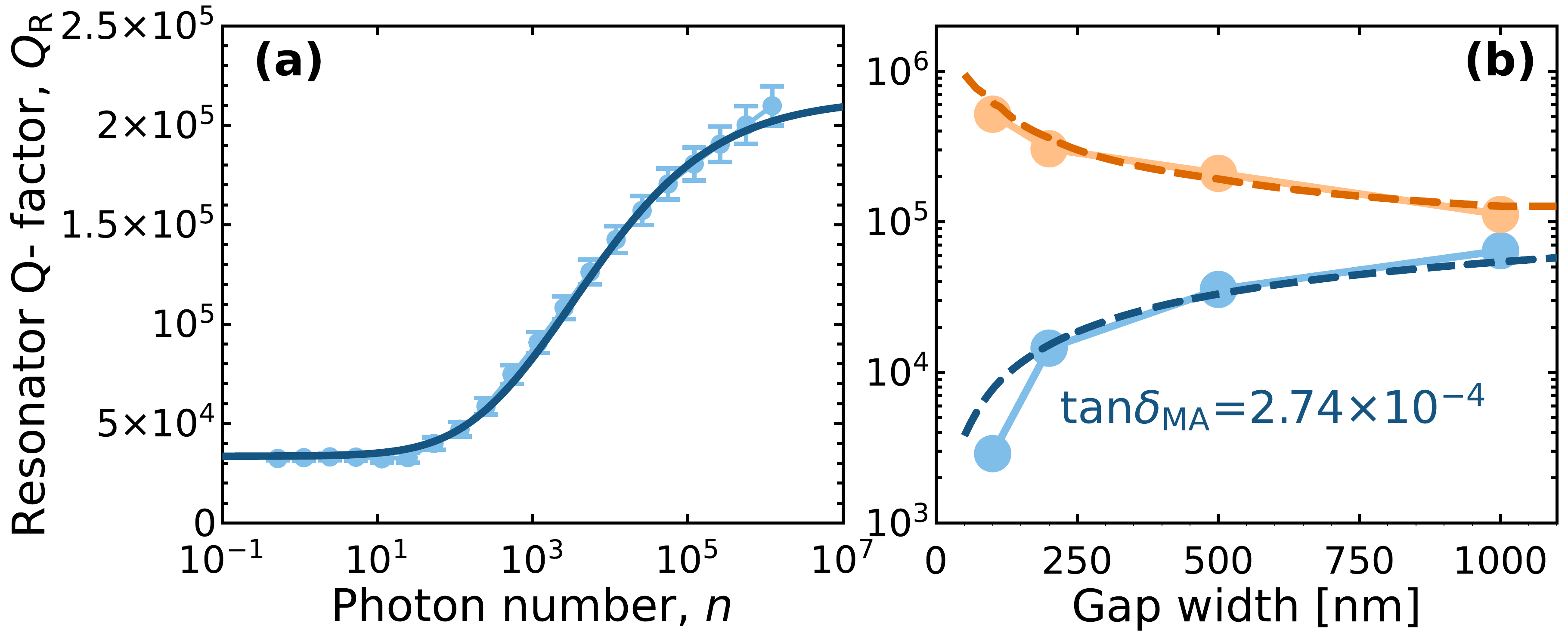}
\caption{\label{fig:resonatorq} 
\textbf{Resonator measurements.} 
(a) Measured internal quality factor $Q_\text{R}$ as a function of the probe power dependent photon number $n$ for a 500\,nm vacuum gap resonator (circles) with the 95\% confidence interval of the joint quadrature fits (error bars) and fit to the TLS model Eq.~\ref{eq:qtot} (blue line) with $n_c=46\pm19$ and $\beta=0.43\pm0.04$. 
(b) Extracted $Q_{\mathrm{R,low}}$ (blue circles) from the fit to Eq.~\ref{eq:qtot} and $Q_{\mathrm{R,high}}$ (orange circles) taken from the highest measured value in the linear regime as a function of vacuum gap width. The blue dashed line is a fit to Eq.~\ref{eq:qtls} yielding $\tan\delta_{\mathrm{MA}}=2.74\times 10^{-4}$ assuming a 3\,nm parasitic layer thickness with a relative permittivity of 10. The orange dashed line is linear fit to the simulated electric field maximum in the vacuum gap $Q_{\mathrm{R,high}}\propto|\vec{E}|$.  
}
\end{figure}

For each resonator, we observe an increase of $Q_{\mathrm{R}}$ for higher probe powers as shown in Fig.~\ref{fig:resonatorq}(a). This is the typical behavior usually assigned to resonant energy absorption by saturable TLS that can be modeled with \cite{gao2008}
\begin{eqnarray} \label{eq:qtot}
\frac{1}{Q_\mathrm{R}}=\frac{1}{Q_{\mathrm{R,low}}}\frac{\tanh\left(\frac{\hbar\omega_\mathrm{R}}{2kT}\right)}{\left(1+\frac{n}{n_c}\right)^{\beta}}+\frac{1}{Q_{\mathrm{R,high}}},
\end{eqnarray}
where $Q_{\mathrm{R,low}}$ and $Q_{\mathrm{R,high}}$ are the limit cases for low and high probe power. All measurements are performed at $T\lesssim  10\,\mathrm{mK}$ where $\tanh{\left(\frac{\hbar\omega_\mathrm{R}}{2kT}\right)} \approx 1$.
The intra-cavity photon number $n$ is calculated from the applied VNA power at the device 
and $n_c$ is the critical photon number required to saturate the TLS. 
The exponent $\beta$ is derived to be $1/2$ for a non-interacting TLS bath and varies between 0.30 and 0.45 to yield very good agreement with the data shown in Fig.~\ref{fig:resonatorq}(a) which suggests TLS interactions in agreement with earlier studies \cite{burnett2017}.

Figure~\ref{fig:resonatorq}(b) shows the extracted $Q_{\mathrm{R,low}}$
and $Q_{\mathrm{R,high}}$ for four resonators with different capacitor gap widths. For non-driven qubit applications the relevant number is $Q_{\mathrm{R,low}}$, which decreases with decreasing capacitor gap width (and its footprint area) as expected since the surface sensitivity to the MA interface is higher as shown in Fig.~\ref{fig:sens}(b). 
This behavior can in general be modeled by the 
weighted sum of the loss tangents of all loss channels
\begin{eqnarray}\label{eq:qtls}
\frac{1}{Q_{\mathrm{TLS}}}=\sum_i p_i \tan\delta_i\approx s_{\mathrm{MA}}\varepsilon_{\mathrm{MA}}t_{\mathrm{MA}}\tan\delta_{\mathrm{MA}},
\end{eqnarray}
where we used $p_{\mathrm{MA}}\gg p_{\mathrm{MS}}, p_{\mathrm{SA}}, p_{\mathrm{Si}}$ to simplify the equation, as justified by the surface sensitivity simulations together with earlier observations that the loss tangents of all interfaces are of the same order of magnitude \cite{wenner2011,wang2015} and Appendix C. 
Using the relative permittivity $\varepsilon_{\mathrm{MA}}=10$ and layer thickness $t_{\mathrm{MA}}=3\,\mathrm{nm}$ together with the simulated values of the sensitivity $s_{\mathrm{MA}}$, as shown in Fig.~\ref{fig:sens}, we fit the measured data of $Q_{\mathrm{R,low}}$ in \ref{fig:resonatorq}(b) (dashed blue line) with very good agreement and extract $\tan\delta_{\mathrm{MA}}=2.74\times 10^{-4}$ at $\approx 9.25$\,GHz. 
This value is in qualitative agreement with the literature for oxidized aluminum exposed to ambient conditions 
for which so far only an upper limit could be extracted due to the limited selectivity of the geometry. Note however that much lower values have been achieved for very low-volume (small junction) aluminum oxide that is not exposed to air \cite{mamin2021}, which is directly relevant for small area JJ dissipation. 

In contrast, $Q_{\mathrm{R,high}}$ shown in Fig.~\ref{fig:sens}(b) is larger for smaller capacitors and very closely follows a dependence proportional to the simulated magnitude of the zero-point electric field ranging from 4 to 22\,V/m in the vacuum gap $Q_{\mathrm{R,high}}\propto|\vec{E}_\text{zpf}|$, see also Appendix B. Here we refer to the 
highest measured value 
before the frequency response of the resonator deviated from Lorentzian behavior due to the meander inductor becoming a nonlinear element at large applied probe powers. 
The MA surface selectivity shown in Fig.~\ref{fig:sens}(b) does not explain this behavior since the TLS are saturated at high drive power.  Similarly, bulk Si losses and the sensitivity to the other interfaces (MS and SA) likely do not play a role due to the small change in energy participation (0.38 to 0.46\,\% for bulk losses) and surface sensitivity for the four different gap sizes. 
The most likely reason for higher measured Q-factors for smaller gaps at high drive powers is that for wider capacitor gaps with lower electric fields we are not able to fully saturate the same fraction of the TLS using a narrow-band coherent drive tone compared to the case of smaller capacitors, see also Appendix B.

The measured resonance frequencies of $\omega_\text{R}/(2\pi)=$ 9.1, 9.5, 9.8, and 8.6\,GHz for the capacitor gaps 100, 200, 500, and 1000\,nm agree very well with the prediction for all gap widths, which makes the capacitor a good candidate to be utilized for compact transmon qubits with a reproducible charging energy $E_\mathrm{C}$. Focussing on qubits we can also exclude potential losses that might arise in the meander inductor and study the single photon limit and TLS interaction in more detail.

\subsection{Qubit measurements}
Vacuum gap transmon qubits are fabricated with the same capacitor design that we use for the resonators. Shunting them with a single JJ with a calibrated Josephson energy $E_\mathrm{J}$ results in a well-defined qubit frequency, see Appendix D for qubit parameters.  As a readout device, we use a lumped element resonator with resonance frequency $f_R\approx 9$\,GHz and total linewidth $\kappa/(2\pi)\approx 0.6$\,MHz patterned on a suspended silicon membrane. Here we use much larger coupling capacitor gaps, as shown in Fig.~\ref{fig:sem},
with an electric field participation of
$10^{-3}$ of the qubit capacitor such that it does not limit the extracted dielectric vacuum gap losses. 

We measure the energy relaxation time $T_1$ using a standard two-tone time-domain technique with a shifting $\pi$-pulse 
that is calibrated with a previously conducted Rabi oscillation measurement. 
We perform a standard dispersive qubit readout \cite{blais2021}
with a very low probe power to stay in the non-demolition regime and avoid measurement induced 
frequency shifts. The qubit population between ground and excited state is extracted from the phase shift of the reflected probe tone. Due to the 
low signal-to-noise ratio in this low power limit without a quantum limited amplifier, we use up to 
10 million averages conducted for up to 10 hours. 

\begin{figure}[t]
\includegraphics[width=1.0\columnwidth]{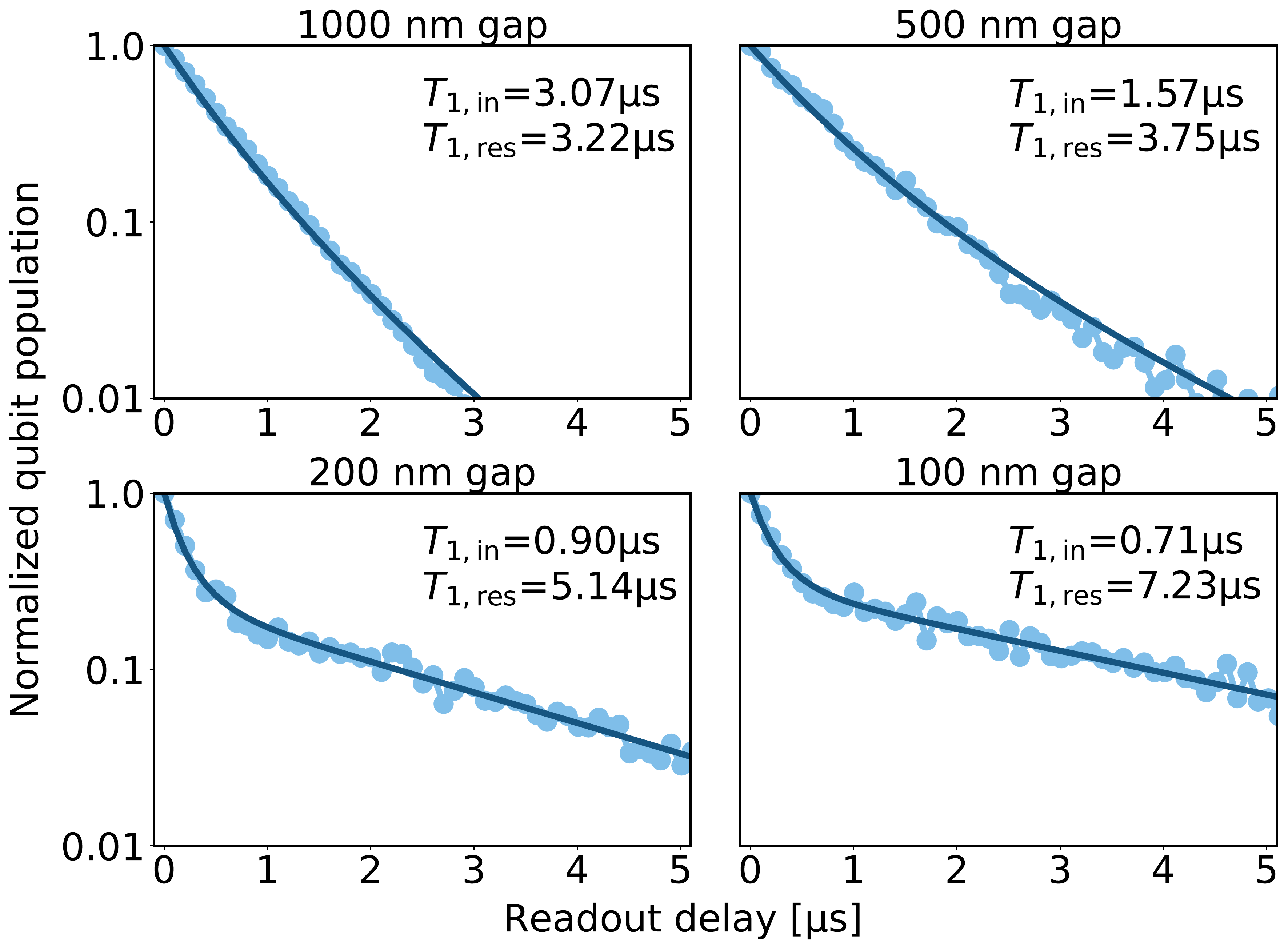}
\caption{\label{fig:t1log} 
\textbf{Energy relaxation measurements.} 
Measured energy relaxation for 4 qubits with different vacuum gap width (circles) fitted with the double exponential model Eq.~\ref{eq:double} (lines) and the extracted $T_1$ times for the initial and residual decay are indicated. 
}
\end{figure}

Figure~\ref{fig:t1log} shows the measured qubit population decay in logarithmic scale for 4 different gap widths ranging from 100 to 1000\,nm. 
In the case of smaller gaps the observed decay exhibits a very pronounced double exponential character.
A similar decay was observed before \cite{pop2014,gustavsson2016} and ascribed to non-equilibrium quasiparticles causing the fast initial decay, followed by a constant slower decay due to residual relaxation channels. 
We use the same phenomenological model \cite{pop2014}
\begin{eqnarray} \label{eq:double}
P(t)=e^{<n_{\mathrm{in}}>(\exp{([-t/T_{1,\mathrm{in}}]-1)})}e^{-t/T_{1,\mathrm{res}}},
\end{eqnarray}
in order to extract the initial 
and the residual 
energy relaxation time constants $T_{1,\mathrm{in}}$ and $T_{1,\mathrm{res}}$ as a result of TLS interactions.
In analogy to the quasiparticle case, here the double exponential behavior is scaled by the normalized concentration of participating TLS $n_{\mathrm{in}}$, which we fit to be between 0.5 to 2.5 as the capacitor gap is increased. 

\begin{figure}[t]
\includegraphics[width=\columnwidth]{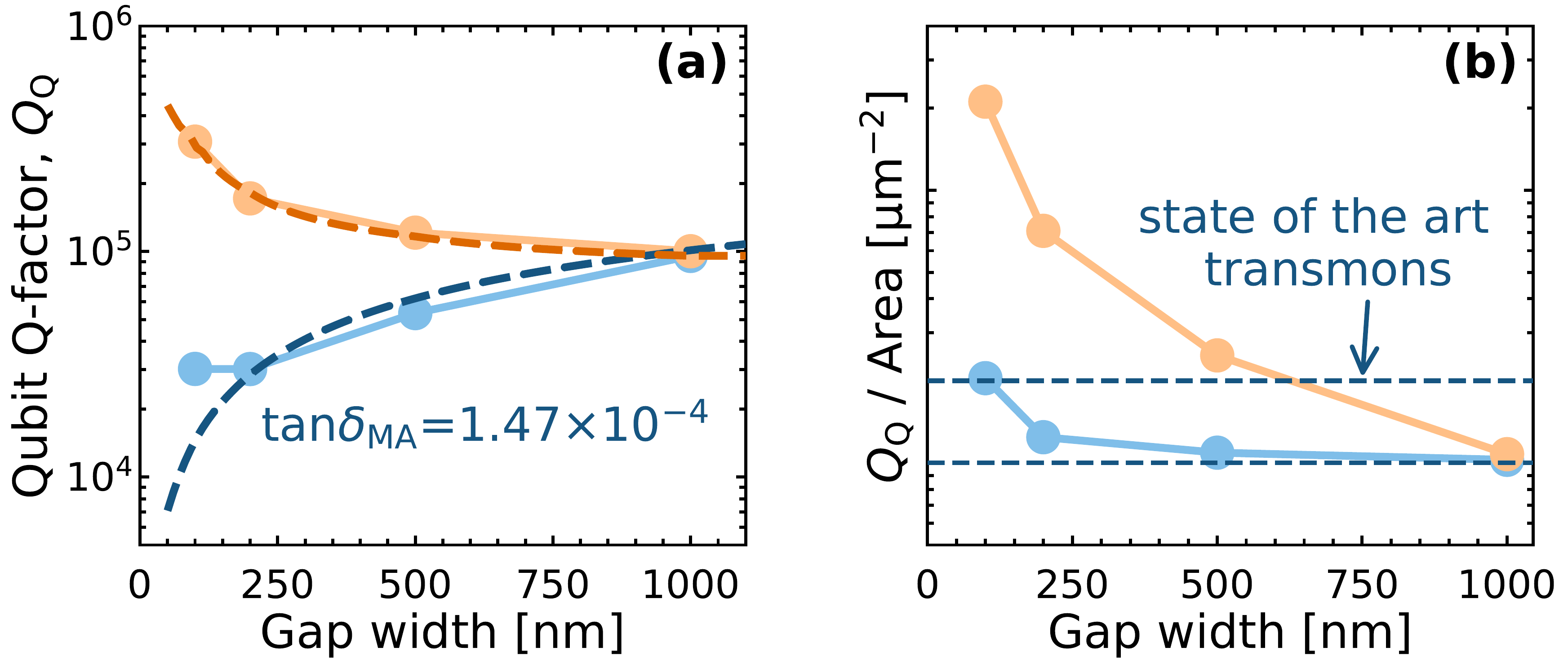}
\caption{\label{fig:qubitq} 
\textbf{Qubit quality factor and integration density.} 
(a) The qubit Q-factors obtained from energy relaxation measurements related to the initial and the residual population decay as a function of capacitor gap width. $Q_{\mathrm{Q,in}}$ (blue circles) corresponds to the initial decay $T_{\mathrm{1,in}}$ which follows the expected trend due to the increased surface sensitivity of smaller gaps 
quantified by Eq.~\ref{eq:qtls} (blue dashed line). $Q_\mathrm{Q,res}$ corresponds to the measured residual decay time $T_{\mathrm{1,res}}$ (orange circles) is found to scale proportional to the magnitude of the electric field in the gap $Q_{\mathrm{Q,res}}\propto|\vec{E}|$ (orange dashed line). 
(b) $Q_{\mathrm{Q,in}}$ (blue circles) and $Q_\mathrm{Q,res}$ (orange circles) normalized by qubit footprint area $A$ as a function of gap width. For comparison, dashed lines indicate state-of-the-art transmon qubit results from Refs.~\cite{arute2019,place2021}.
}
\end{figure}

Each time constant can be assigned to the respective qubit quality factor $Q_{\mathrm{Q,in}}=2\pi f_\text{g-e}T_{1,\mathrm{in}}$ and $Q_{\mathrm{Q,res}}=2\pi f_\text{g-e} T_{1,\mathrm{res}}$, with $f_\text{g-e}$ the ground to first excited state qubit transition frequency. In Fig.~\ref{fig:qubitq}(a) we show that $Q_{\mathrm{Q,in}}$ is dropping with decreasing capacitor gap width (blue circles) which can be explained by the limitation due TLS in analogy to the low photon number Q-factor $Q_{\mathrm{R,low}}$ of the resonator. 
Using the simulated values of $s_{\mathrm{MA}}$ for the specific capacitor geometry and the expected properties of the parasitic layer ($t_{\mathrm{MA}}=3\,\mathrm{nm}, \varepsilon_{\mathrm{MA}}=10$) to fit the data (blue dashed line) with Eq.~\ref{eq:qtls} we extract the value of the loss tangent for the parasitic layer on the metal surface of the qubit capacitor $\tan{\delta}_{\mathrm{MA}}=1.47\times10^{-4}$ at $\approx 5.4$\,GHz. This is a factor 1.9
lower than the value obtained from resonator measurements, which is to a large part due to the lower qubit frequency by a factor $\approx 1.7$ taking the average qubit and resonator frequencies. Another contribution could be parasitic
losses in the stray capacitance of the resonator meander inductor, which we didn't include in our simulations. We also find that both a saturation pulse, or a $\pi$-pulse train experiment yields an improvement of up to a factor 2, which we attribute to TLS saturation and coherence rather than non-equilibrium quasiparticles \cite{gustavsson2016}, see Appendix E. 
  
Interestingly, the Q-factor assigned to the residual decay $Q_{\mathrm{Q,res}}$ has the opposite trend, see orange circles in Fig.~\ref{fig:qubitq}(a), with a qualitatively very similar dependence to the high photon number resonator Q-factor $Q_{\mathrm{R,high}}$, and also follows the dependence $Q_{\mathrm{Q,res}}\propto|\vec{E}|$ with very good agreement to the measurements, see orange dashed line in Fig.~\ref{fig:qubitq}(a). 
It was recently shown \cite{tai2021,spiecker2022} that the TLS ensemble can have a high effective coherence, which in the case of large zero-point electric fields can cause coherent energy exchange between a small TLS ensemble with a higher probability to find the qubit in the excited state after its intrinsic $T_1$ time determined by the initial decay, see also Appendix B. In fact, vacuum Rabi splitting between a resonator mode and a single TLS has been observed before in monolithic \cite{sarabi2015,sarabi2016} and vacuum gap capacitors \cite{fink2016} and strong coupling between qubits and TLS located in the JJ have even been controlled \cite{grabovskij2012} and used as a microscopic quantum memory \cite{neeley2008}.
We conclude that the larger coupling to a smaller capacitor increases the coupling to a smaller ensemble of coherent TLS 
which then appears as a slower residual decay of the average qubit population. In a few measurements we even found indications of an increased excited state probability at long time scales as numerically modeled in Ref.~\cite{tersoff2021}.
Alternatively, it might also be possible that the higher sensitivity to other surfaces and bulk losses in the case of less confined electric fields would lead to a longer residual decay but our simulations indicate that those losses are too small to have an appreciable impact, see also Appendix C. 

To quantify the qubits' integration possibility and compare its performance with larger footprint designs we define the ratio of qubit Q-factor per footprint area $Q/A$ as a figure of merit. If we consider the Q-factor determined by the initial fast qubit decay, see blue circles in Fig.~\ref{fig:qubitq}(b), we get $Q_\mathrm{Q,in}/A$ of slightly above 20\,$\upmu$m$^{-2}$ for the smallest qubit. This is similar to the state-of-the-art planar transmon qubits. For comparison, the big pads of tantalum transmon qubits \cite{place2021} with a $T_1$ up to 300\,$\upmu$s exhibit also a value $Q/A\approx20\,\upmu$m$^{-2}$ and for the Google chip demonstrating quantum supremacy \citep{arute2019} it is $Q/A\approx10\,\upmu$m$^{-2}$. Looking at $Q_\mathrm{Q,res}/A$ (orange circles in Fig.~\ref{fig:qubitq}b), which roughly follows the TLS saturated limit in case of resonators, gives a sense that this figure of merit could be improved by orders of magnitude using lower-loss superconductor surfaces. This strongly motivates the careful study of metal surface treatments and the use of lower loss superconductors. 

\section{Conclusion}
In this work we present an alternative to the monolithic approach \cite{mamin2021,antony2021, wang2022} to reducing the physical size of superconducting qubits. 
Vacuum capacitors are insensitive to substrate losses, a design principle that could also further improve the coherence of large footprint circuits. 
The extremely compact qubit size down to $\approx 40\,\upmu\text{m}^2$ studied in this work and the low effective dielectric permittivity $\epsilon_\text{eff}\approx 1$ drastically reduces radiative coupling \cite{sage2011,rafferty2021} and thus also the related radiation loss and stray light induced charge noise \cite{pan2022}. Moreover, since the qubit circuit is suspended on a silicon nano-membrane it is also expected to be highly protected from bulk phonons, e.g.~those originating from high energy radiation absorbed in the thick handle wafer substrate \cite{mcewen2022}, in particular when combined with full phononic bandgap shielding on patterned SOI slabs \cite{safavi2010}.

We use simulations and measurements to confirm that the remaining dominant loss mechanism is related to TLS in the metal-air interface. This makes compact vacuum gap transmons extremely selective and sensitive probes for superconductor surface losses.
We utilize this property to precisely quantify the dielectric surface losses of electron-beam evaporated aluminum of $\varepsilon_\text{MA} t_\text{MA} \tan{\delta} = 10\times 3\,\text{nm} \times 1.47\times10^{-4}$ at $\approx 5.4$\,GHz. The resulting TLS-limited 
Q-factor defined by the initial population decay of qubits follows the expected surface sensitivity simulations with a Q-factor over footprint area of up to $\gtrsim 20\,\upmu$m$^{-2}$, which is on par with the longest $T_1$ transmon qubits \cite{arute2019,place2021}. 
Moreover, for small qubits with high zero-point electric fields we observe a double exponential energy relaxation 
with a residual decay mechanism that correlates with the TLS saturated Q-factor of $LC$ resonators. The dependence on the gap width and electric field is consistent with an increased coupling to high coherence TLS \cite{tai2021,spiecker2022}. 

Better understanding and suppressing TLS losses is a major goal in the field and 
could be achieved for instance by actively engineering the noise spectrum  \cite{you2022}, avoiding sample oxidation, through superconductor surface treatments, or by using different materials with a much lower surface loss tangent, such as TiN \cite{sage2011,sandberg2012,woods2019}, NbTiN \cite{barends2010} or Ta 
\cite{place2021,wang2021}. 
Using these materials with the presented vacuum gap geometry 
combined with high efficiency readout and control of the TLS bath \cite{spiecker2022} would shine more light on the TLS loss properties of various surfaces and advance the development of a high integration density and low-loss superconducting circuit architecture needed for large-scale error corrected quantum processors.

The data and code used to produce the figures in this manuscript will be made available on the Zenodo repository.

\section*{ACKNOWLEDGMENTS}
This work was supported by the Austrian Science Fund (FWF) through BeyondC (F7105), the European Research Council under grant agreement number 758053 (ERC StG QUNNECT) and a NOMIS foundation research grant. M.Z. was the recipient of a SAIA scholarship, E.R. of a DOC fellowship of the Austrian Academy of Sciences and M.P. of a P\"{o}ttinger scholarship at IST Austria. S.B. acknowledges support from Marie Sk{\l}odowska Curie fellowship number 707438 (MSC-IF SUPEREOM). The authors acknowledge 
support from the ISTA Nanofabrication Facility. 


\section*{Appendix A: Device fabrication}\label{fab}
\begin{figure*}[t]
\centering
\includegraphics[width=1\textwidth]{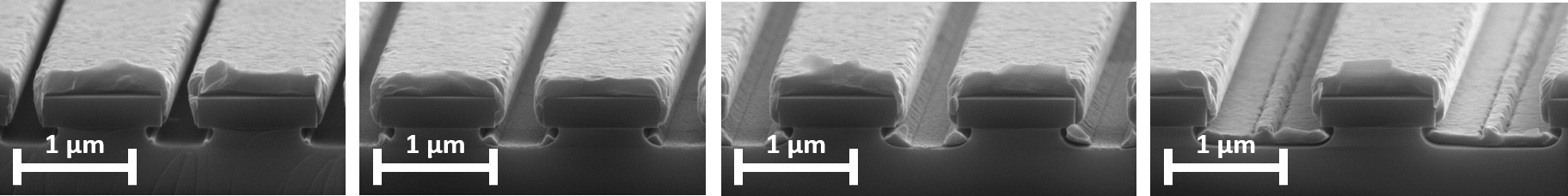}
\caption{\label{fig:allgaps_sem} 
\textbf{Vacuum gap capacitors SEM images.} Isometric view of (from left) 100\,nm, 200\,nm, 500\,nm, and 1000\,nm wide capacitor gaps of a cleaved sacrificial sample after angle evaporation with aluminum in the UHV system. In the cross-section we see the SiO$_2$ supporting the silicon beams. For the real device is this layer is removed as the last step by  HF vapor resulting in a fully suspended vacuum gap finger capacitor. The leftover metal in the gaps relocates to the silicon handle wafer which is separated by 3\,$\upmu$m.}
\end{figure*}

We fabricate the devices for this study using commercially available silicon-on-insulator (SOI) wafers diced to 1$\times$1\,cm$^2$ sized chips. The wafer consists of a 220\,nm thin high-resistivity silicon device layer ($>$3\,k$\Omega$cm, (100) FZ, P-type Boron) separated by 3\,$\upmu$m thermally grown silicon dioxide (SiO$_2$) from the 725\,$\upmu$m thick silicon handle wafer ($>$5\,k$\Omega$cm, (100) CZ, P-type Boron). The first step was to remove the dielectric in the gaps of the capacitor and form 200\,nm diameter through-holes which ensures access of the hydrofluoric acid (HF) vapor in the last process step. After thorough initial cleaning with acetone, isopropanol, and buffered hydrofluoric acid (BHF) we spin and bake the electron beam sensitive resist CSAR and expose the structures with a high-resolution electron-beam lithography system (Raith EBPG5150) at 100\,kV. After the development of the exposed resist the silicon device layer is dry-etched with inductively coupled plasma reactive ion etching (ICP-RIE) generated in a mixture of SF$_6$ and C$_4$F$_8$ gases. After that, we use BHF to create a 100\,nm deep undercut that prevents unwanted circuit shortcuts and ensures that HF vapor can access the SiO$_2$ in the last process step. Then we use angle evaporation in an ultra-high vacuum (UHV) Plassys MEB550S2 electron-beam evaporator. We evaporate 80\,nm thick aluminum from a 45\textdegree angle from both sides each in an interleaved fashion to realize a homogeneous coating of the silicon beam top and side walls. The cross-section of a cleaved sample right after this process step is shown in Fig.~\ref{fig:allgaps_sem} depicting all investigated capacitor gaps.

\begin{figure}[t]
\includegraphics[width=1\columnwidth]{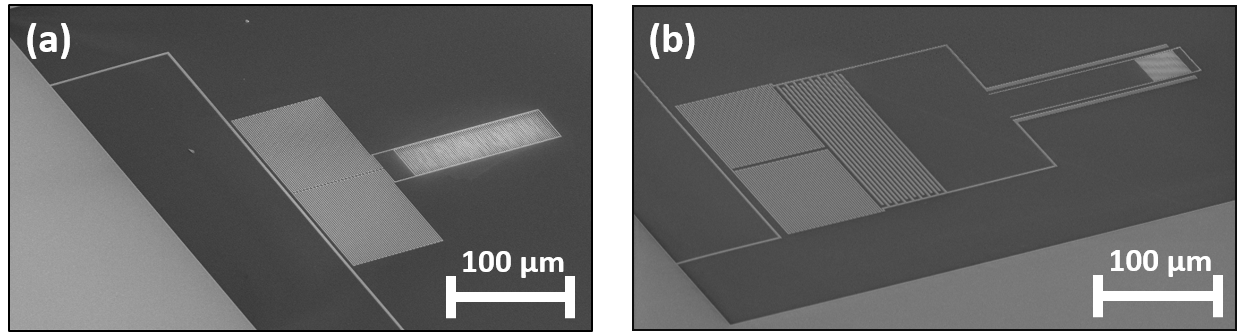}
\caption{\label{fig:alldev_sem} 
\textbf{Fabricated vacuum gap capacitor devices.} 
(a) Lumped element resonator with 500\,nm wide capacitor gap. 
(b) Transmon qubit with a 100\,nm wide vacuum gap capacitor capacitively coupled to a readout resonator.}
\end{figure}

In the next step we pattern the aluminum layer (meander inductor, coplanar waveguide launcher, and bond pads) using e-beam lithography and 
ICP dry etching with 
a mixture of Cl$_2$ and BCl$_3$ gasses. Then we fabricate the JJ with the standard Dolan bridge technique using a double layer of 600\,nm MAA and 300\,nm PMMA resist. After e-beam lithography and development we use the same UHV system to evaporate 60\,nm and 120\,nm of aluminum from a $\pm$25\textdegree angle followed by a lift-off process in hot (80\textdegree\,C) N-Methyl-2-Pyrrolidone (NMP). To ensure a good galvanic connection between JJ and capacitor electrodes we make an additional 200\,nm thick aluminum patch layer with an in-situ ion gunning step right before the evaporation. In the last process step we etch the SiO$_2$ BOX layer in HF vapor for $\sim$7\,hours releasing the suspended membranes in the region defined by the etched capacitor gaps and the silicon through-holes that were etched in the first process step. Figure~\ref{fig:alldev_sem} shows a finished LC resonator and qubit device each.

After the sample fabrication, we glue the chip to a copper printed circuit board (PCB), and wire-bond the on-chip bond-pads to the coplanar waveguides on the PCB which is equipped with surface mount soldered MMPX coaxial connectors. The PCB with the chip is enclosed in a copper box designed to prevents parasitic box modes.

\section*{Appendix B: Electric field dependence of the Q-factor}
\begin{figure}[t]
\centering
\includegraphics[width=0.3\textwidth]{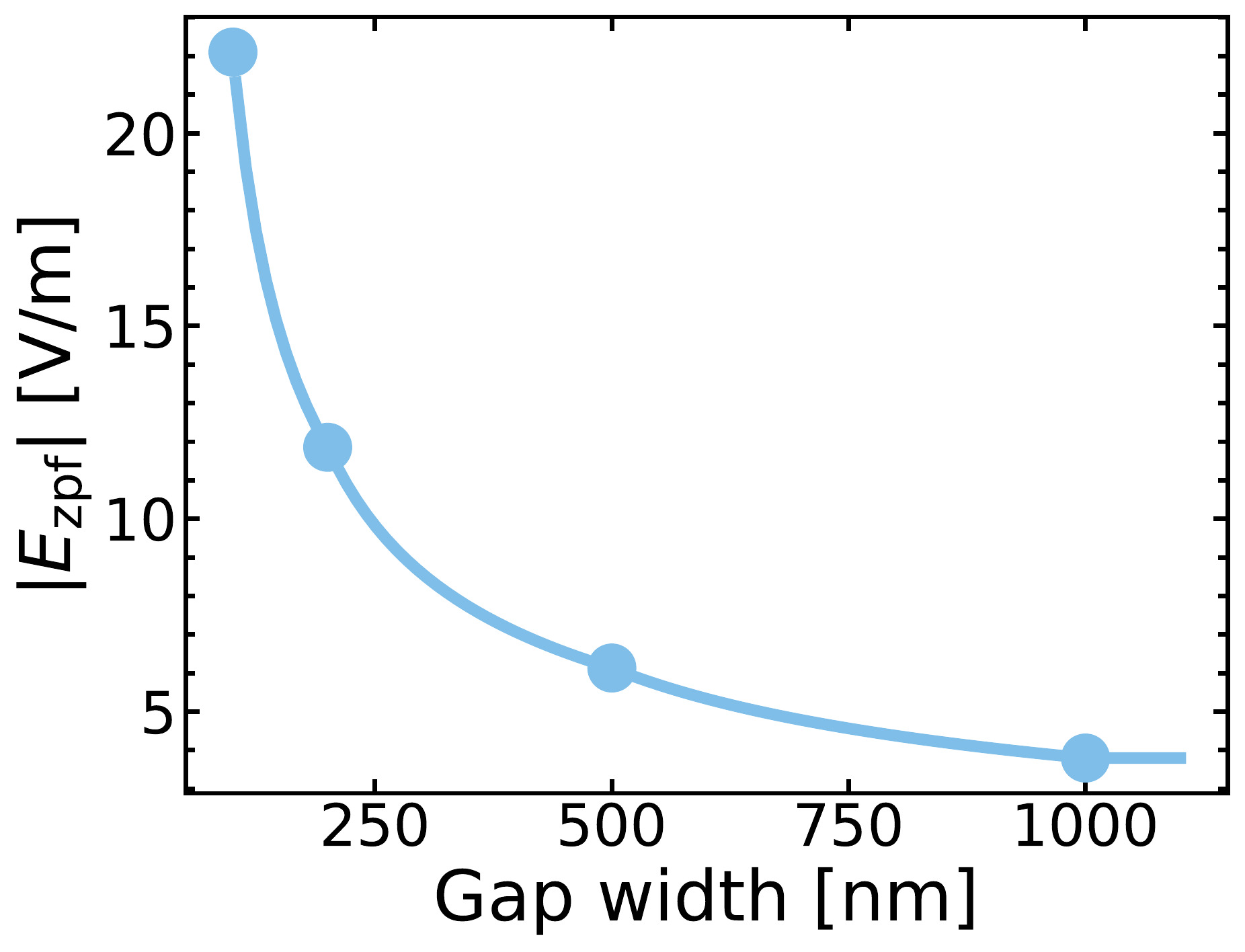}
\caption{\label{fig:zpf} \textbf{Simulated zero-point electric field.} Simulated magnitude of the electric field as a function of gap width. 
The simulation is done for an applied voltage of 1\,V to the capacitor electrodes and then normalized to the zero-point fluctuations corresponding to the energy $\frac{1}{2}\hbar\omega$. Circles highlight the capacitor gap width of the devices which were fabricated.}
\end{figure}

In the main text, we show that the Q-factor of lumped element resonators using a vacuum gap capacitor measured at high applied power $Q_{\mathrm{R,high}}$, as well as the Q-factor of the vacuum gap transmon qubits related to the residual decay of the qubit population $Q_{\mathrm{Q,res}}$, both follow a linear dependence on the magnitude of zero-point electric field 
in the capacitor $\propto|\vec{E}|$, shown in Fig.~\ref{fig:zpf} as a function of vacuum gap capacitor width. 

A recent study \cite{tai2021} showed a detailed theoretical and experimental analysis of the TLS and quasi-particle (QP) influence on the Q-factor of coplanar waveguide resonators in the low power and temperature limit. An observed increase of the Q-factor with decreasing temperature below 50\,mK is ascribed to an increased coherence time of the TLS. The formula describing this behavior can be approximated in the low-temperature limit studied here ($k_BT\ll$ TLS energy level splitting) with the simple equation
\begin{eqnarray} \label{eq:QTLS0}
Q_{\mathrm{TLS}}=Q_{\mathrm{TLS,0}}\sqrt{1+\frac{\xi}{T} |\vec{E}|^2},
\end{eqnarray}
where $\xi$ is the power and temperature-independent parameter describing the TLS properties, $T$ is the temperature, and $Q_{\mathrm{TLS,0}}$ is the lower Q-factor limit. Fitting the $Q_{\mathrm{R,high}}$ and $Q_{\mathrm{Q,res}}$ as a function of gap size with Eq.~\ref{eq:QTLS0} yields the fit parameters $\xi$ and $Q_{\mathrm{TLS,0}}$ summarized in Tab.~\ref{tab:table}. Since in our case the second term under the square root is $\gg1$ this is equivalent to the fits $\propto|\vec{E}|$ shown in the main text. For completeness we also include the fitted $\tan\delta$ extracted from fitting $Q_{\mathrm{R,low}}$ and $Q_{\mathrm{Q,in}}$ as function of gap size in Tab.~\ref{tab:table}. 

\begin{table}[h]
\centering
\begin{tabular}{| C{2cm} | C{2cm} | C{2cm} | C{2cm} |}
\hline
device & tan$\delta$ & $\xi$ [$\mathrm{Km^2/V^2}$] & $Q_{\mathrm{TLS,0}}$\\
\colrule
resonators & 2.74$\times 10^{-4}$ & 3$\times 10^{-3}$ & 5.5 $\times 10^4$\\
qubits & 1.47$\times 10^{-4}$ & 0.3$\times 10^{-3}$ & 8 $\times 10^4$\\
\hline
\end{tabular}
\caption{\label{tab:table} \textbf{TLS material parameters.} Parameters extracted from fitting the gap dependent high power Q-factors obtained from resonators and the Q-factors from the residual qubit decay, respectively, using Eq.~\ref{eq:QTLS0}.}
\end{table}

The extracted fitting parameters for strongly driven resonators are $Q_{\mathrm{TLS,0}}=5.5\times 10^4$ and $\xi=3\times 10^{-3} \,\mathrm{Km^2/V^2}$, which is in very good agreement with this value for glassy TLS systems found in \cite{tai2021}. One argument why this model might also be relevant to the high power ($\sim10^5$ photons) regime is that the applied narrow-band pump may not be able to saturate all the TLS in the relevant frequency band before the resonator switches to the nonlinear regime.
The residual TLSs would then be less dissipative for smaller capacitor gaps by the same principle as in \cite{tai2021} since the electric field per excitation is stronger. 

For the residual qubit relaxation data we obtain the fit parameters $Q_{\mathrm{TLS,0}}=8\times 10^4$ and $\xi=0.3\times 10^{-3}$ which is ten times lower than in the high power resonator case. The largest uncertainty in the extracted $\xi$ values lies in the assumed temperature of $T$, which we fixed to be the measured mixing chamber temperature of 10\,mK. However, the electromagnetic temperature of the cavity mode is expected to be significantly higher, typically in the range from 50 to 90\,mK for the implemented shielding and attenuation, and might even increase for the strongly driven resonator data $Q_{\mathrm{R,high}}$.  

\section*{Appendix C: MS, SA and bulk loss tangents}
To extract the loss contributions of the MS and SA surfaces, we fabricate a transmon qubit with a much bigger finger capacitor (5\,$\upmu$m finger width, 5\,$\upmu$m gap width, 400$\times$400\,$\upmu$m$^2$ area) on a suspended silicon membrane but without the vacuum gap, as shown in Fig.~\ref{fig:density}(b). This capacitor has a much higher surface sensitivity of the electric field to the substrate interfaces SA and MS compared to a vacuum gap capacitor, see Fig.~\ref{fig:sens}(a), such that none of the three loss channels can be neglected.
For this qubit, we measure a single exponential energy relaxation decay with a $T_1=2.41\,\upmu s$. 
Using the perviously determined value for $\tan{\delta}_{\text{MA}}$ based on vacuum gap qubits and assuming that the MS and SA interfaces have the same loss tangent, thickness of 3\,nm and relative permittivity $\varepsilon=10$, we extract $\tan{\delta}_{\text{MS, SA}}\approx 3.0\times 10^{-4}$, just slightly higher than the MA interface. However to quantify this value with better precision a more detailed study involving various geometries with a bigger variation of the electric field surface sensitivity to these interfaces would be needed.

For the vacuum gap devices we neglect the dielectric losses in bulk silicon since its participation is only 0.4\%, which gives, assuming even the pessimistic limit for the silicon loss tangent $\tan{\delta}_{\text{Si}}=1\times 10^{-6}$, an upper Q-factor limit of $\sim10^9$, which is significantly above the measured values of up to $6\times10^5$ in this study. While it is theoretically possible that the SOI device layer has a much increased loss tangent compared to typically used bulk silicon this would have led to a lower $T_1$ in the MS, SA test device discussed earlier, with a silicon substrate participation of order 80\,\%.

\section*{Appendix D: Qubit parameters}\label{qubits}
The parameters of the 4 measured qubits are given in Tab.~\ref{tab:qubits}.
\begin{table}[h]
\centering
\begin{tabular}{| C{1cm} | C{1.2cm} | C{1cm} | C{1cm} | C{1cm} | C{1cm} | C{1cm} |}
\hline
gap (nm) & area ($\upmu$m$^2$) & $f_\text{g-e}$ (GHz) & $E_\text{C}/h$ (MHz) & $E_\text{J}/h$ (GHz) & $T_{1,\text{in}}$ ($\upmu$s)& $T_{1,\text{res}}$ ($\upmu$s)\\
\colrule
100 & 39$\times$36 & 5.94 & 461 & 11.11 & 0.71 & 7.23\\
200 & 39$\times$62 & 5.33 & 389 & 10.51 & 0.9 & 5.14\\
500 & 39$\times$125 & 5.41 & 362 & 11.51 & 1.57 & 3.75\\
1000 & 39$\times$239 & 4.96 & 342 & 10.28 & 3.07 & 3.22\\
\hline
\end{tabular}
\caption{\label{tab:qubits} \textbf{Qubit parameters.} Capacitor gap size, qubit capacitor area, ground to excited state frequency, charging energy, Josephson energy, initial and residual energy relaxation times for the four measured qubits are reported.}
\end{table}

\section*{Appendix E: TLS pumping}
To improve the coherence properties further we pump the qubit with a $\pi$-pulse train \cite{gustavsson2016} as well as a long resonant saturation pulse right before the $T_1$ measurement sequence. Both yield a close to factor 2 improvement of the measured initial $T_1$ but only when the $\pi$-pulses in the train are separated by short time scales such as $\sim 100$\,ns. 
This supports the statement that the qubit coherence is limited by TLS rather than non-equilibrium quasiparticles because TLS can get saturated with a long coherent pulse, which is not expected for quasiparticles. The highest measured initial relaxation time for the smallest qubit reached $T_1=1.23\,\upmu$s, which improves the figure of merit for integration density $Q/A\approx 40\,\upmu\text{m}^{-2}$.

\bibliography{vgq_submission}

\end{document}